\documentclass{elsart}
\usepackage{amssymb}
\usepackage{amsmath}
\usepackage{epsfig}
\usepackage{graphicx}

\journal{\ Int. J. Engineering Science, vol. \textbf{32}, 10 (1994)
pp. 1553-1560\quad\quad\quad
\quad\quad\quad\quad\quad\quad\quad\quad\quad\quad\quad\quad\quad}
\input{tcilatex}
\begin{document}

\begin{frontmatter}
\title{VIBRATIONS OF LIQUID DROPS\\ \ IN FILM BOILING PHENOMENA
\\ (\emph{the mathematical model})}

\author{Pierre Casal  \ \emph{\small and\ }}  \author[Francia] {Henri
Gouin} \ead{ henri.gouin@univ-cezanne.fr \\ Corresponding author}
\address[Francia]{University of Aix-Marseille \& U.M.R. C.N.R.S.
6181\\
 Av. Escadrille Normandie-Niemen, Box 322, 13397 Marseille Cedex 20, France.  }

{\small \texttt{\emph{In memory of Professor Pierre Casal}}}
\begin{abstract}
Flattened liquid drops poured on a very hot surface evaporate quite
slowly and float on a film of their own vapour. In the cavities of a
surface, an unusual type of vibrational motions occurs.  Large
vibrations take place and different forms of dynamic drops are
possible. They form elliptic patterns with two lobes or hypotrochoid
patterns with three lobes or more. The lobes are turning relatively
to the hot surface. We present a model of vibrating motions of the
drops. Frequencies of the vibrations are calculated regarding the
number of lobes. The computations agree with experimental forms
obtained in [1] by  Holter and  Glasscock.
\end{abstract}

\end{frontmatter}

\section{INTRODUCTION}

Water splashed on a moderately hot metallic flat plate spreads out, comes to
the boil and then quickly evaporates. It is not the same when the metal is
very hot: water remains cool, breaks up into many drops that roll, bound and
are thrown on all sides. Such drops are mentioned in the literature as being
in "spheroidal state". Most of the people have noticed these phenomena in
their childhood when they stared at rolling drops on a very hot oven.
\newline
Leidenfrost first experimented the process in 1756: a little water poured on
a red hot spoon, does not damp the spoon and takes the same shape as
mercury. A strong motion of vapour lying between liquid and metal supports
liquid masses and causes fast vibrating motion in the liquid bulk. This is
the film boiling or Leidenfrost phenomenon [2]. It is well known since
before the time of Faraday and was the subject of numerous studies during
the nineteenth century [3]. \newline
One of the interests in the Leidenfrost phenomenon rises from an explosive
behaviour associated with non-equilibrium of pressures. This affects all the
industrial sectors where hot temperatures are used (for example in metal
industry: quenching of metals [4], etc). This generates the unsteadiness of
distillation plants in the petroleum industry and many accidents in nuclear
engineering [5]. \newline
Indeed, the first surprising effect was previously described. The liquid
drops are floating on a vapour film when they are placed on hot plates. This
state is usually explained as the fact that when the temperature of the wall
is greater than a value depending both on the fluid and the state of the
surface, the exchanges of heat are small corresponding to the formation of a
thin film of vapour isolating drops from the wall. In such conditions,
liquid drops maintain during a time of the order of minutes and usual
descriptions present the film of vapour keeping liquid drops well below
boiling. Observations of the drops give phenomena by translational and
rotational motions. Photographic and stroboscopic tools have been used to
have descriptions of the experiments, but the effects can be seen by naked
eye. These vibrations lead to beautiful patterns that appear as solid
geometric figures. Such a phenomenon is qualitatively very well described in
the paper by Holter and Glasscock [1]. \newline
The different models do not give an interpretation of vibrating motions in
the liquid bulk of drops on hot plates. It seems that classical treatments
of vibrations (like in Lamb [6]) do not cover the present situation.
Contrary to first impression, we determine that the present situation
complicated by the gravity does not take into account heat gradients, vapour
flows or more subtle effects. \newline
For the analytic approach, we consider a horizontal surface with a slight
cavity invariant by rotation around its vertical symmetry axis. The drop is
assumed not quite big enough to slip out of the cavity. Experiments describe
such drops as flattened spheroids in the vertical plane. If the mean
curvature radius of the surface is large compared to the size of the
flattened drops, we assume that the motion of the liquid is approximately a
motion in the horizontal plane. We consider plane motions of an
incompressible liquid submitted to a convenient potential due to the effects
of curvature of the surface. We investigate only \emph{rotations} of drops
and not random motions occurring by translation if the surface is really
flat plate. \newline
Due to the film of vapour, we assume that motions are frictionless and the
very small viscosity effect is neglected. The motions are \emph{not} small
vibrations. The form of drops in vibration are hypotrochoids [7] with
different numbers of sides or lobes. The case of only two lobes corresponds
to an elliptic form of drops. The results of computation favourably compare
to experiments. It is possible to evaluate the angular velocity of
vibrations. It depends on the acceleration of gravity, the curvature of the
surface and the number of lobes. Times of rotation for these spheroidal
drops are computed. They seem to be in accord with simple observations [8].

\section{A CLASS OF PLANE MOTIONS FOR INCOMPRESSIBLE PERFECT FLUIDS}

\subsection{Motions of a fluid}

The motion of a continuous flow can be represented by a surjective
differential mapping
\begin{equation}
\zeta \rightarrow \chi =\Phi (\zeta )\,,
\end{equation}%
where $\zeta =(t,\xi )$ belongs to $\mathcal{W}$, an open set in the
time-space occupied by the fluid between time $t_{1}$ and time
$t_{2}$. The position in the reference space $\mathcal{D}_{o}$ is
denoted by $\chi $; its position at time $t$ in $\mathcal{D}_{t}$ is
denoted by $\xi $ [9]. We assume that distinct points of the
continuous fluid remain distinct throughout the entire motion. At
$t$ fixed, transformation (1) possesses an inverse.

In order to describe a particular case of plane motions of a fluid let us
introduce relatively to fixed Cartesian plane systems of coordinates, the
inverse mapping
\begin{equation}
Z\rightarrow z=\varphi (t,Z).
\end{equation}%
The complex number $Z=X+iY$ is such that $\chi =(X,Y)\in
\mathcal{R}^{2}$ represents Lagrangian coordinates in the reference
space $\mathcal{D}_{o}$\,. The complex number $z=x+iy$ with $\xi
=(x,y)$ $\in \mathcal{R}^{2}$ represents Eulerian coordinates in the
Euclidean plane $\mathcal{D}_{t}$. Let us notice that $(X,Y)$ is not
necessarily a spatial position of a particle. Then, $(X,Y)$ are
\emph{material }coordinates.

In polar form $Z=re^{i\theta }$. Here we call $(r,\theta )$ the
reference position. Now, we write $(r,\theta )\in \mathcal{D}_{o}$
and in polar form, transformation (2) is written
\begin{equation}
(r,\theta )\in \mathcal{D}_{o}\rightarrow z=\varphi _{t}(r,\theta
)\in \mathcal{D}_{t}\,.
\end{equation}%
Let us consider a particular transformation (2) of the form
\begin{equation}
z=Z+\overline{f(Z)}\,e^{i\omega t},
\end{equation}%
where $\omega $ is a real constant and the function $f=p+iq$ \ is an
analytic function of $Z$ with $p$ is the real part and $q$ is the
imaginary part of $f$. \newline Components of the velocity ${\mathbf
V}$ are $({\dot{x}},{\dot{y}}) $ and $\dot{z}=i\omega
\overline{f(Z)}\,e^{i\omega t}$.
\newline
Components of the acceleration are $(\ddot{x}, \ddot{y})$ and
$\ddot{z}=-\omega ^{2}\,\overline{f(Z)}\, e^{i\omega t}$. \newline
Let us denote $F(Z)$ a primitive function of $f(Z$). At $t$ fixed,
relation (4) yields
\begin{equation*}
e^{-i\omega t}f(Z)dz=e^{-i\omega t}f(Z)dZ+(p+iq)(dp-idq)
\end{equation*}%
or
\begin{equation*}
e^{-i\omega t}f(Z)dz=e^{-i\omega t}dF(Z)+ \frac{1}{2}\,
d(p^{2}+q^{2})+i(qdp-pdq)
\end{equation*}%
and
\begin{equation*}
\mathcal{R[}e^{-i\omega t}f(Z)dz]=d\Phi
\end{equation*}%
with
\begin{equation}
\Phi =\mathcal{R[}e^{-i\omega
t}f(Z)]+\frac{1}{2}f(Z)\overline{f(Z)}\,.
\end{equation}%
($\mathcal{R}$ denotes the real part of a complex number).

Consequences of transformation (4) are:
\begin{equation*}
\qquad (a)\qquad \qquad \dot  {x}dy-\dot{y}dx-i(
\dot{x}dx+ \dot{y}dy)=-i\,\overline{\dot {z}}%
dz=-\omega \,e^{-i\omega t}f(Z)dz
\end{equation*}%
and consequently $ \dot{x}dy-\dot{y}dx=-\,\omega \ d\Phi
$\,.\newline Then
\begin{equation}
div\,{\mathbf V}=0
\end{equation}%
and $-\omega\, \Phi $ is the stream function.
\begin{equation*}
\qquad (b)\qquad \qquad \ddot{x} dx+\ddot{y}
dy+i(\ddot{x}dy-\ddot{y}dx)=\overline{ \ddot{z}}dz=-\omega
^{2}\,e^{-i\omega t}f(Z)dz
\end{equation*}%
and consequently ${\ddot{x}}dx+{\ddot{y}}dy$ $  =-\omega
^{2}d\Phi\,.$ \newline Then,
\begin{equation}
{\mathbf\Gamma} =-\omega ^{2}\func{grad}\ \Phi\,.
\end{equation}%
Equation (7) shows that the acceleration ${\mathbf\Gamma} $ is the
derivative of a potential.

\subsection{Pressure of these plane motions}

In the case of plane motion of an incompressible perfect fluid, the
equation of motion yields
\begin{equation}
{\mathbf\Gamma} =-\frac{1}{\rho }\, \func{grad}\, p-\func{grad}\, W,
\end{equation}%
where $\rho $ is the constant volumic mass, $p$ is the pressure and
$W$ is the extraneous force potential. \newline From equation (7) we
get
\begin{equation}
\frac{p}{\rho }=\omega ^{2}\Phi -W.
\end{equation}%
Relation (9) yields the value of the pressure field as a function of $\Phi $
and $W$.

\section{PARTICULAR CLASS OF PLANE MOTIONS}

For a reference set $\mathcal{D}_{o}$ such that
\begin{equation}
\mathcal{D}_{o}\equiv\left\{ Z = r\,e^{i\theta}\ \ {\rm with}\ \ \big( r\in \left[ 0,r_{o}\right] ,\,\theta \in %
\left[ 0,2\pi \right] \big)\right\} .
\end{equation}%
Let us consider two particular cases of motions defined by a function f such
that:
\begin{equation*}
\begin{array}{l}
\begin{array}{llllllllll}
(a) &  &  & f(Z)=\displaystyle\frac{Z^{n}}{a^{n-1}}, &  & n\in N, &  & n> 1,
&  & a\in \mathcal{R}^{+\ast }%
\end{array}
\\
\begin{array}{llllllllll}
(b) &  &  & f(Z)=\lambda\, Z, &  & \lambda \in\, ]0,1[ &  &  &  &
\end{array}%
\end{array}%
\end{equation*}

\textit{3.1\quad Case (a)}
\begin{equation}
Z\in \mathcal{D}_{o}\ \longrightarrow\ z=\
Z+\frac{\overline{Z^{n}}}{a^{n-1}}\, e^{i\omega t}\in
\mathcal{D}_{t}
\end{equation}
is the representation of the motion. \\ In parametric
representation, relation (11) yields:
\begin{equation}
\begin{array}{l}
x=r\cos \theta + \displaystyle\frac{r^{n}}{a^{n-1}}\cos (\omega t-n\theta )
\\
y=r\sin \theta\, +\displaystyle\frac{r^{n}}{a^{n-1}}\sin (\omega t-n\theta )%
\end{array}%
\end{equation}%
with $(r,\theta )\in \mathcal{D}_{o}.$

The trajectory of a particle with reference position $(r,\theta )$ is a
circle $(\mathcal{C}_{r,\theta })$. In the system coordinates, the centre of
$(\mathcal{C}_{r,\theta })$ is $(r\cos \theta ,r\sin \theta )$ and the
radius is $R=\displaystyle r^{n}/a^{n-1}$ . The particle associated $%
(r,\theta )$ moves on the circle $(\mathcal{C}_{r,\theta })$ with the
angular velocity $\omega $. \newline
At time $t$, particles whose reference positions in $\mathcal{D}_{o}$ are on
the circle with center $O$ and radius $r$ move in the physical space $%
\mathcal{D}_{t}$ on a hypotrochoid curve $\mathfrak{E}_{t}(r)$. Curves $%
\mathfrak{E}_{t}(r)$ are obtained as circular disks of radius
$\displaystyle\rho =r/n$ rolling internally inside a fixed circle of
radius $r+\rho $.
\newline Two different points of $\mathcal{D}_{o}$ correspond to two
different points of $\mathcal{D}_{t}$ and the hypotrochoid has no
double point. Due to
\begin{equation*}
\det \displaystyle%
\begin{vmatrix}
\displaystyle\frac{\partial x}{\partial r} & \displaystyle\frac{\partial x}{%
\partial \theta } \\
\displaystyle\frac{\partial y}{\partial r} & \displaystyle\frac{\partial y}{%
\partial \theta }%
\end{vmatrix}%
=r(1-n^{2}\,\frac{r^{2n-2}}{a^{2n-2}}),
\end{equation*}%
the condition for the hypotrochoid has no double point is $\displaystyle%
\left( \frac{r}{a}\right) ^{n-1}<\displaystyle\frac{1}{n}$ and the set $%
\mathcal{D}_{o}$ of material variables verify
\begin{equation}
\left( \frac{r_{o}}{a}\right) ^{n-1}<\frac{1}{n}\,.
\end{equation}%
Consequently, $\mathfrak{E}_{t}(r_{o})$ is the free boundary of the
fluid and the length $a$ verifies
\begin{equation*}
a>r_{o}\sqrt[n-1]{n}\,.
\end{equation*}%
One verifies that
\begin{equation*}
\varphi _{t}(r,\theta )=\varphi _{o}(r,\theta -\Omega t)e^{i\Omega
t}
\end{equation*}%
with
\begin{equation*}
\Omega =\frac{\omega }{n+1}\,.
\end{equation*}%
The hypotrochoids $\mathfrak{E}_{t}(r)$ where $r\in \lbrack 0,r_{o}]$ and
specially the free boundary $\mathfrak{E}_{t}(r_{o})$ of the fluid, are
turning relatively to the hot plate with the angular velocity $\Omega =%
\displaystyle\frac{\omega }{n+1}$ .

\textit{3.2\quad Case (b)}
\begin{equation}
Z\in \mathcal{D}_{o}\rightarrow z=\ Z+\lambda \overline{Z}e^{i\omega
t}\in \mathcal{D}_{t}
\end{equation}%
is the representation of the motion. \newline In parametric
representation, relation (14) yields:
\begin{equation}
\begin{array}{l}
x=r\cos \theta +\lambda r\cos (\omega t-\theta ) \\
y=r\sin \theta +\lambda r\sin (\omega t-\theta )%
\end{array}%
\end{equation}%
with $(r,\theta )\in \mathcal{D}_{o}$. Contrary to case $(a)$, no
limitation is imposed on $r_{o}$\,. \newline At time $t$, the
particles whose reference positions in $\mathcal{D}_{o}$ are on the
circle centered on $O$ and radius $r$ move in the physical space on
an elliptic curve $\mathfrak{E}_{t}(r)$. Ellipses
$\mathfrak{E}_{t}(r)$ have semi-axes of length $r(1+\lambda )$ and
$r(1-\lambda )$. They are
turning relatively to the hot plate with the angular velocity $\displaystyle%
\Omega = \omega/2 $.

\section{ MOTIONS OF A PERFECT FLUID IN THE BOTTOM OF A LARGE CAVITY.}

With a convenient Euclidean frame $Oxyz$, the equation of the hot surface is
\begin{equation}
\begin{array}{lllll}
z=f(r) &  & \mathrm{with} &  & r^{2}=x^{2}+y^{2} .
\end{array}%
\end{equation}%
Function $f$ is a $C^{2}$-function and $f(0)=f^{\prime }(0)=0$. The
vertical direction is $Oz$. Equation (16) yields
\begin{equation}
\begin{array}{lllll}
z=\displaystyle\frac{f^{\prime \prime
}(0)}{2}\,r^{2}+r^{3}\varepsilon (r) & & \mathrm{with} &  & \
\lim_{r\rightarrow 0} \, \varepsilon (r)=0\,.
\end{array}%
\end{equation}%
For a surface whose meridian curve is of small curvature, we use the
equivalent approximation
\begin{equation}
z=\frac{f^{\prime\prime }(0)}{2}\,r^{2} .
\end{equation}%
The potential of gravity forces is
\begin{equation}
\begin{array}{lllll}
W=k(x^{2}+y^{2}) &  & \rm{with} &  & \displaystyle k=g\frac{f^{\prime\prime }(0)}{%
2}\, .
\end{array}%
\end{equation}%
Relation (18) is used in the following computation: the mean
curvature of the surface at $(0,0,0)$ is
$R=\displaystyle\frac{g}{2k}$  and
\begin{equation}
k=\frac{g}{2R}\,.
\end{equation}%
Let us consider the two cases $(a)$ and $(b)$:

\quad\quad \emph{In case} $(a)$
\begin{equation*}
f(Z)=\frac{Z^{n}}{a^{n-1}}\quad \Longrightarrow \quad
F(Z)=\frac{Z^{n+1}}{(n+1)\,a^{n-1}}\,.
\end{equation*}%
We deduce the stream function,
\begin{equation*}
\Phi =\frac{r^{2n}}{2a^{2n-2}}+\frac{r^{n+1}}{(n+1)a^{n-1}}\cos
[\omega t-(n+1)\theta ]\,.
\end{equation*}%
For the potential (19) we deduce from relation (12) the expression
\begin{equation*}
W=k\left[ r^{2}+\frac{r^{2n}}{a^{2n-2}}+2\frac{r^{n+1}}{a^{n-1}}\cos
\left(\omega t-(n+1)\theta \right)\right] .
\end{equation*}%
Equation (10) yields
\begin{eqnarray*}
&&\quad\quad\quad\quad \frac{p}{\rho } = \omega ^{2}[\frac{r^{2n}}{2a^{2n-2}}%
+\frac{r^{n+1}}{(n+1)a^{n-1}}\cos (\omega t-(n+1)\theta )] \\
&&\quad\quad\quad\quad -k[r^{2}+\frac{r^{2n}}{a^{2n-2}}+2\frac{r^{n+1}}{%
a^{n-1}}\cos (\omega t-(n+1)\theta )]+C^{te}.
\end{eqnarray*}%
In the case $\omega ^{2}=2(n+1)k$ and for a pressure $p_{o}$ on the
free boundary associated with $r=r_{o}$, we obtain
\begin{equation}
\frac{p}{\rho }=\frac{p_{o}}{\rho }+\frac{\omega ^2}{2(n+1)}(r_{o}^2 -r^2)-%
\frac{n}{2(n+1)}\,\omega ^2\, \frac{r_{o}^{2n}-r^{2n}}{a^{2n-2}}\,.
\end{equation}
\quad \quad \emph{In the case }$(b)$
\begin{equation*}
\begin{array}{lll}
F(Z)=\displaystyle\frac{\lambda }{2}\,Z^{2}, &  & \Phi =\displaystyle\frac{%
\lambda ^{2}}{2}\,r^{2}+\frac{\lambda }{2}\,r^{2}\cos (\omega t-2\theta )%
\end{array}%
\end{equation*}%
and
\begin{equation*}
W=kr^{2}[1+2\lambda \cos (\omega t-2\theta )+\lambda ^{2}]\,.
\end{equation*}%
Equation (12) yields
\begin{equation*}
\frac{p}{\rho }=r^{2}\omega ^{2}\,[\frac{\lambda ^{2}}{2}+\frac{\lambda }{2}%
\cos (\omega t-2\theta )]-kr^{2}\,[1+2\lambda \cos (\omega t-2\theta
)+\lambda ^{2}]+C^{te}.
\end{equation*}%
In the case $\omega ^{2}=4k$ and for a pressure $p_{o}$ on the free
boundary, we obtain
\begin{equation*}
\frac{p}{\rho }=\frac{p_{o}}{\rho }+\frac{\omega
^{2}}{4}\,(1-\lambda ^{2})(r_{o}^{2}-r^{2}).
\end{equation*}%
In the two cases, equation (20) yields $\omega ^{2}=\displaystyle
 (n+1)g/{R} $ and fixes the angular rotation of the drop
\begin{equation}
\Omega ^{2}=\frac{g}{(n+1)R}\,.
\end{equation}%
Consequently, following the different values of $n$, different modes
are possible.
\newline The period of the rotation of the vibrating drop is given
by
\begin{equation}
T=2\pi \sqrt{\frac{(n+1)R}{g}}\,.
\end{equation}%
In fact, after an interval of time $T_{n+1}=\displaystyle
{T}/{(n+1)}$, the
drop is identical to its previous position and $\displaystyle\nu _{n+1}=%
 {1}/T_{n+1}$ is the apparent frequency of changes.
\begin{equation}
\nu _{n+1}=\frac{1}{2\pi }\sqrt{\frac{(n+1)g}{R}}\,.
\end{equation}

\section{ COMPUTATION AND GRAPHS OF THE MOTION}

Experiments were performed with a very simple apparatus in [1]. No
data were performed, but it seems possible to investigate the above
frequencies. Some apparent frequencies of the drop vibrations are
calculated in Table 1.
\newline
We have drawn different cases of vibrating drops from two lobes
(basic mode) to eight lobes. Drops are turning around their center
of symmetry and we have drawn different positions of a drop with two
and three lobes (Fig. 1). In the case of three-lobed modes, the
trajectory of a particle is represented as a circle for a particular
drop. The graphs have the same forms than the ones  given in [1].
Only the case of two lobes present some discrepancy with the
interpretation given in [1]. In [1], the middle of the drop may be,
in some cases, squeezed by comparison with the shape given here.
Depending on the value of $a$ and $n$, the drop can be like
\emph{vibrating polygons} of any type (Fig. 2). The two orientations
are similar by computation; we notice also it is possible that the
drop reverses the direction of rotation of the motion.
\begin{table}[tbp]
\centering
$%
\begin{tabular}{|c|c|c|c|c|c|}
\hline
\multicolumn{1}{||c|}{$n\ \rightarrow $} & 1 & 2 & 3 & 4 &
\multicolumn{1}{|c||}{5} \\ \hline
\multicolumn{1}{||c|}{$\downarrow \ R$} & $\nu _{2}$ & $\nu _{3}$ & $\nu
_{4} $ & $\nu _{5}$ & \multicolumn{1}{|c||}{$\nu _{6}$} \\ \hline\hline
10 & 2.23 & 2.73 & 3.15 & 3.52 & 3.86 \\ \hline
20 & 1.58 & 1.93 & 2.23 & 2.49 & 2.73 \\ \hline
30 & 1.29 & 1.58 & 1.82 & 2.04 & 2.23 \\ \hline
40 & 1.11 & 1.37 & 1.58 & 1.76 & 1.93 \\ \hline
50 & 1.00 & 1.22 & 1.41 & 1.58 & 1.73 \\ \hline
\end{tabular}
$%
\caption{ \ Some apparent frequencies $\protect\nu_{n+1}$ (in Hertz)
of drops as function of the mean curvature radius of the surface (in
cm) and the number n+1 of lobes of the drop (g=981
cm/s$^2$).\newline } \label{TableKey}
\end{table}

\bigskip

\begin{figure}[tbp]
%[th]
\par
\begin{center}
\epsfig{file={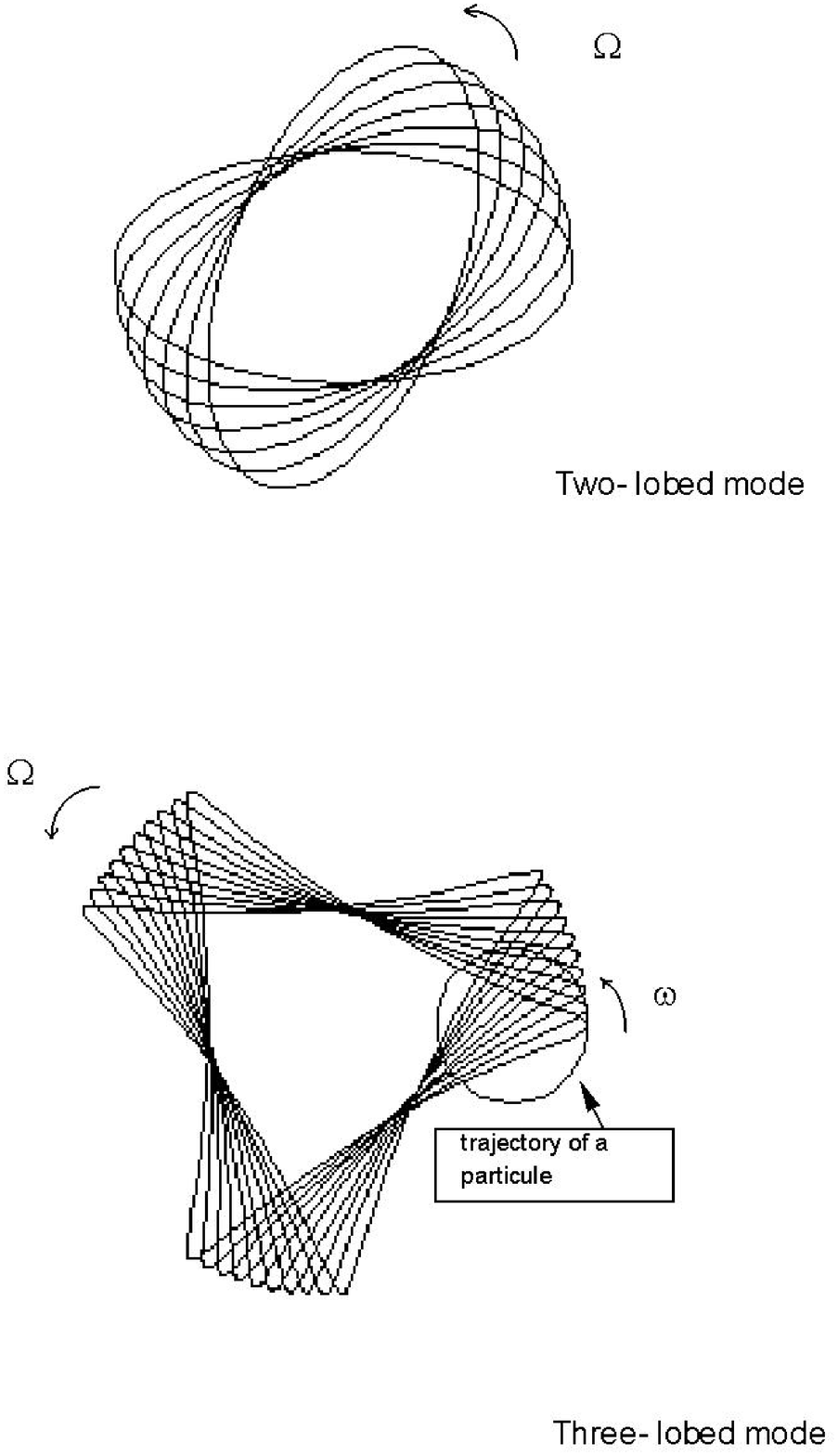}, width = 0.90\linewidth}
\end{center}
\caption{In the case of two-lobed mode or three-lobed mode, the
drops are turning around their centre of symmetry. The trajectory of
a particle is a circle represented in the case of a three-lobed
mode.} \label{fig1}
\end{figure}

\begin{figure}[tbp]
%[th]
\par
\begin{center}
\epsfig{file={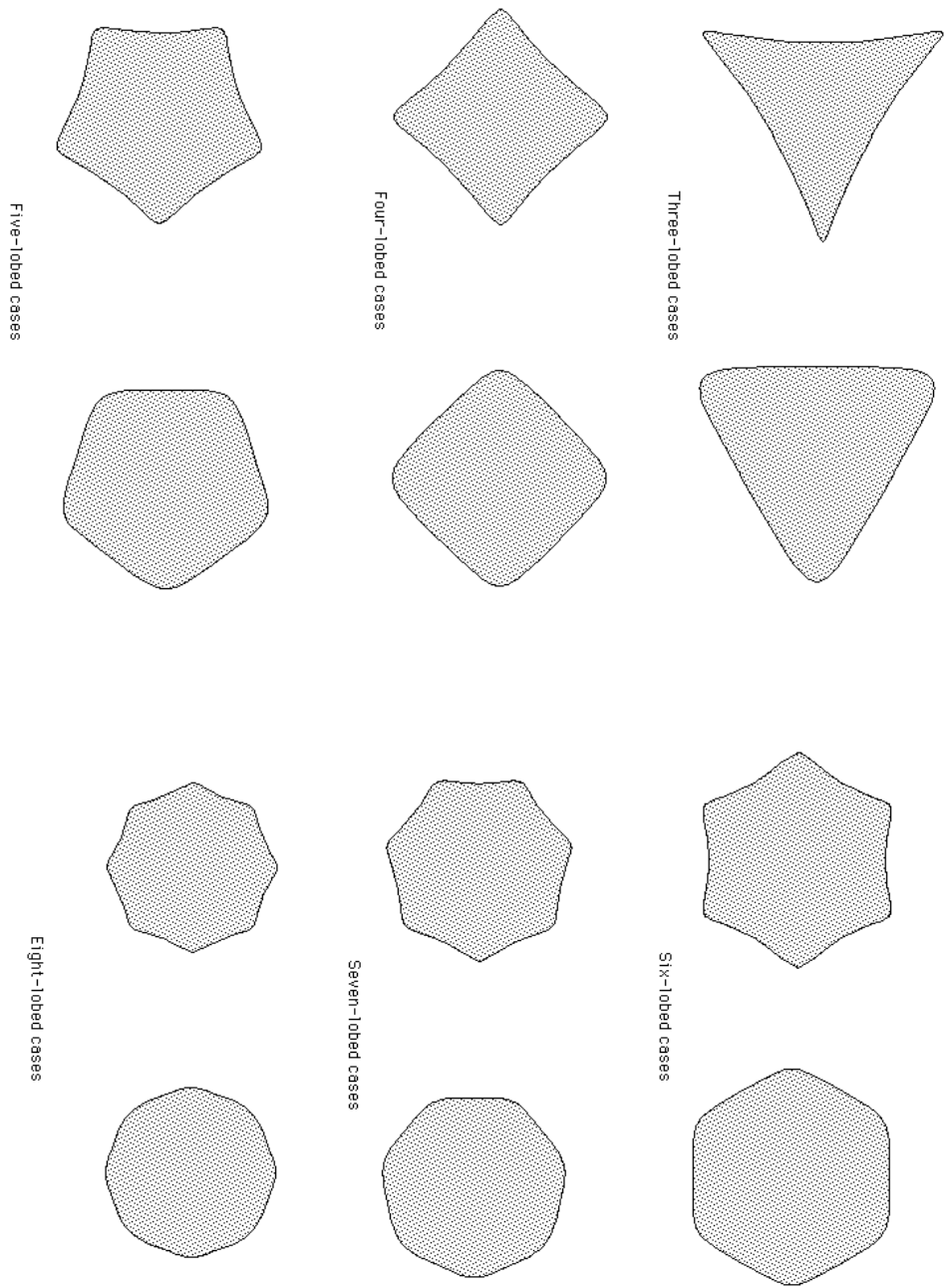}, width = 1.1 \linewidth}
\end{center}
\caption{Two possible forms of drop are presented in cases of three-lobed to
eight-lobed modes. In many cases the vibrating drops are approximately
polygons which may have any of three to \emph{n} sides.}
\label{fig2}
\end{figure}

\end{document}